# Extension of the operation mode of back-streaming white neutron beam at the China spallation neutron source

Xiaolong Gao[a,b,c] Hantao Jing[b,c,*] Chungui Duan[a] Binbin Tian[b,c,d] Yang Li[b,c] Xiaoyun Yang[b,c,e] Jilei Sun[b,c] Weiling Huang[b,c]

[a] *Hebei Normal University, Shijiazhuang 10094, China*

[b] *Spallation Neutron Source Science Center, Dongguan 523803, China*

[c] *Institute of High Energy Physics, Chinese Academy of Sciences, Beijing 100049, China*

[d] *School of Energy and Power Engineering, Xi'an Jiaotong University, Xi'an 710049, China*

[e] *Guangxi Normal University, Guilin, 541004, China*

**Abstract**

The back-streaming white neutron source (Back-n) is a comprehensive neutron experimental platform for nuclear data measurement, nuclear astrophysics, neutron irradiation, detector calibration, etc. In order to meet a variety of experimental requirements, an attempt of various combinations of collimators has been made for the current collimation system. The basic parameters such as beam flux and beam spot characteristics under different beam operation modes have been studied. According to the change of the CSNS proton beam operation mode, the influence of different proton beam spots on the Back-n beam is also studied. The study finds that some beam modes with new collimator combinations can meet the experimental requirements and greatly shorten the experimental time. In addition, we have found a parameterized formula for the uniform beam spot of the high-power proton beam based on the beam monitor data, which provides a more accurate proton-beam source term for future white neutron beam physics and application researches.

## 1. Introduction

A white neutron beam with a wide energy spectrum has a wide range of applications including nuclear data measurement [1, 2], nuclear astrophysics [3], detector calibration, radiation damage, and atmospheric neutron irradiation [4]. The energy for each neutron can be determined by the time-of-flight method. The white neutron experimental facilities that carry out research on high-precision nuclear data measurements in the world include GELINA [5], n-TOF [6], and LANSCE [7], and radiation effects include ANITA [8], TRIUMF [9], PNPI [10], RCNP [11], and CHIPIR [12], etc. The back-streaming white neutron source is one of the important neutron beamlines on the China Spallation Neutron Source (CSNS). It directly guides a white neutron beam with a wide-energy spectrum from the back-streaming direction of the spallation target. It is the only white neutron beamline currently on service in China. As shown in Figure 1, the Back-n beamline is a medium-length white neutron beamline. Thanks to a high-power accelerator at CSNS, its fluxes can surpass $8\times10^6$ n/cm$^2$/s in the 56-meter and 76-meter experimental halls [13].

The early Back-n research [14] aimed at nuclear data measurements. Since the (n, γ) reaction cross-section measurements have higher requirements for the neutron beam spot size, beam halo range, experimental background and other parameters, the three collimating apertures in the collimation system are using fixed settings. With the increase of user experimental types, especially some experiments without the requirement for a low experimental background such as total cross-section measurements, fission cross-section measurements, cross-section measurements of light-charged particle emitting, and researches of radiation effect, more flexible switching of neutron beam intensity and beam spot size is required. It is necessary to extend the beam operating mode of Back-n to meet the experimental requirements of multiple types of research.

In addition, the Back-n neutron beamline is guided from the proton-incident surface of the spallation target. Compared with the neutron beams guided from the side, the Back-n neutron flux and energy spectrum are more sensitive to the proton beam spot distribution on the target. The parameters of CSNS accelerator had been changed at different stages such as nominal design, commissioning, beam-power increase, and full beam-power operation. Therefore, in this paper we study the influence of different proton beam modes on the Back-n neutron beam. We carry out a parameterized study on the rectangular uniform beam spot according to the measurement data under the full beam-power operation.

In Section 2, a brief introduction to the existing collimation system is given. We also study the neutron beam intensity and beam spot of the Back-n with different combinations of collimating apertures of the collimation system. In Section 3, the beam profile measured by the multi-wire monitor in front of the CSNS target is given and parameterized using a combined function based on the error function. In Section 4, we analyze the influence of different proton beam spots on the Back-n neutron beam. Finally, a summary is given.

## 2. Back-n neutron beamline

The Back-n neutron beamline is one of the neutron beamlines at CSNS. The neutrons fly to the remote experimental halls along the back-streaming direction concerning the proton beam. The sample positions of the two experimental halls are 56 meters and 77 meters away from the target respectively. The endstation 1 (ES#1) with higher neutron flux is closer and the endstation 2 with high resolution (ES#2) is farther as shown in Figure 1.

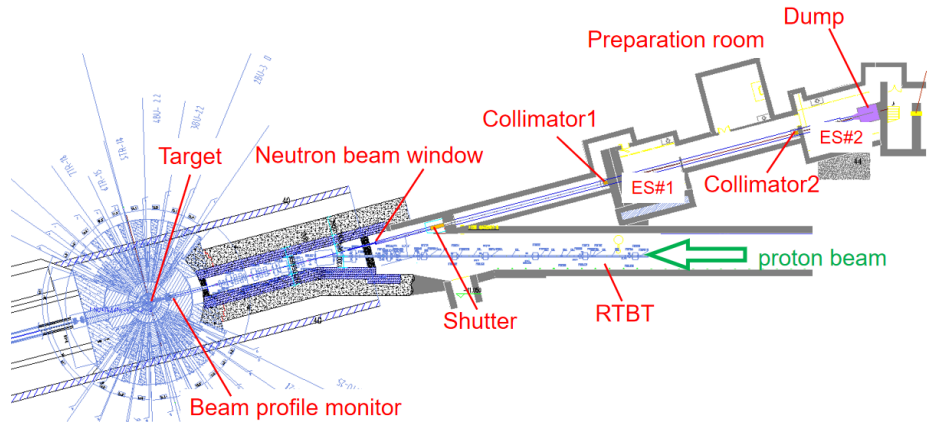

Fig. 1: Layout of Back-n.

The CSNS employs an intense proton beam with an energy of 1.6 GeV and a power of 100 kW to bombard a tungsten target to produce neutrons. For the spallation target, a sliced tungsten target structure is adopted and the tungsten slices are cooled by water. The size of the tungsten target is 160 mm (Width) ×60 mm (Height), and the total thickness is about 650 mm [15, 16], and they are encapsulated in a stainless target container. There are different types of neutron moderators above and below the target container, and they are wrapped by a reflector. The size of the proton-incident surface of the target container is 170 mm (Width) ×70 mm (Height), which is also the emitting surface of the Back-n neutron beam.

Many nuclear data measurement experiments require a high-precision and low experimental background condition. Especially a high-resolution $BaF_2$ detector array system with a $4\pi$ solid angle coverage [17] requires a uniform size of Ø30 mm of the neutron beam spot, which the beam halo range is of less than Ø54 mm, and the experimental gamma and neutron background of less than seven orders of magnitude compared with the flux of the neutron beam. Therefore, the collimation system of the Back-n neutron beamline needs to meet these requirements. The collimation system adopts a three-cascade collimation scheme [16]. The primary collimator (Shutter, also as a shutter) is placed at 31 meters from the target; The main collimator (Coll#1) is placed at 50 meters from the target; The beam halo collimator (Coll#2) is placed at 70 meters as shown in Figure 1. In addition to the small beam spot of Ø30 mm, there are other beam spots such as Ø60 mm and 90 mm×90 mm to meet the requirements of other nuclear data measurement experiments. The aperture of each collimator in the collimation system is shown in Table 1 and each collimator can independently switch the collimating aperture through the remote control system of the driving motor.

Table 1: Positions and apertures of collimators on the beamline of Back-n.

| Beam spot at ES#2 | Shutter (mm) /31 m | Coll#1 (mm) /51 m | Coll#2 (mm) /70 m |
|---|---|---|---|
| Φ20 mm | Φ3 | Φ15 | Φ40 |
| Φ30 mm | Φ12 | Φ15 | Φ40 |
| Φ60 mm | Φ50 | Φ50 | Φ58 |
| 90 mm×90 mm | 78×62 | 76×76 | 90×90 |

## 3. Flux and energy spectrum

Before the completion of the CSNS facility, we had carried out the physical design of the Back-n neutron beamline and given the corresponding intensity and energy spectrum at endstations on the condition of three sets of nominal beam spots. Due to the distribution uncertainty of the proton beam spot, a uniform distribution of the proton beam spot is employed [16]. In 2018, the CSNS facility was completed and started commissioning. From then on, the accelerator began to gradually increase the beam power on the target from 20 kW to 80 kW. Before the power was up to 80 kW, the beam spot distribution of the proton beam on the target was not uniform. According to the analysis from the fluorescence imaging system in front of the target [18], the proton beam spot was an approximately Gaussian distribution. The full width at half maximum (FWHM) of the proton distribution in the horizontal and vertical directions is about 88mm and 33mm respectively. At present, the CSNS accelerator is operating at a beam power of 100 kW, and the proton beam spot is homogenized by a special magnetic lattice before hitting the target.

On the other hand, for many experiments without high-precision requirements, it is necessary to increase the neutron beam intensity to improve the experimental efficiency. We also need to study the flexible combination of the three sets of collimating apertures in the collimation system to meet the diverse needs of users.

In FLUKA, the characteristic parameters of the white neutron beam at the neutron beam window (24m from the target) are obtained by bombarding the spallation target with a proton beam. The energy spectrum distribution is shown in Figure 2. It can be found that the neutron energy spectrum is very wide, spanning ten orders of magnitude, ranging from thermal neutron energy to several hundred MeV and with a peak around 1 MeV.

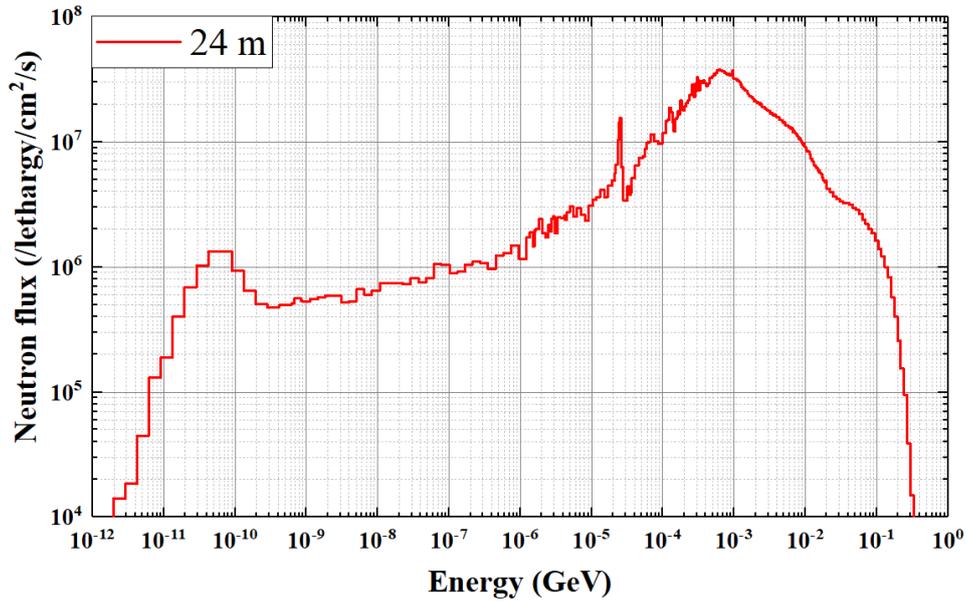

Fig. 2: Energy spectrum of the Back-n neutron beam at 24m from the target.

A Φ20-mm aperture on the shutter is set in consideration of the experiment of extremely low flux and is not a normal operation mode. The approximate Gaussian distribution is used to calculate the beam spot sizes and fluxes for the 27 combination modes, as shown in Table 2. The aperture of the first collimator of the collimation system directly determines the flux in the two endstations. Since the rectangular target is used, an asymmetric elliptical neutron beam spot shape appears in the ES#2. For some cases of arbitrary combination, the uniformity of the beam spot also deteriorates.

Table 2: Beam spot sizes and fluxes in two endstations for the 27 combination modes. The three collimating apertures are given from the second to third columns. The beam spots size and fluxes of two endstations are given from the fifth column to the eighth column respectively. Symbols of Ø, φ, and # indicate circular, elliptical, and rectangular beam spot respectively.

| No. | Shutter (mm) | Coll#1 (mm) | Coll#2 (mm) | ES#1 Beam spot (mm) | ES#2 beam spot (mm) | ES#1 flux (n/cm²/s) | ES#2 flux (n/cm²/s) |
|---|---|---|---|---|---|---|---|

| | | | | | | |
|---|---|---|---|---|---|---|
| 1 | Ø12 | Ø15 | Ø40 | φ18×18 | φ30×30 | 1.67E6 | 6.41E5 |
| 2 | Ø50 | Ø50 | Ø58 | φ54×54 | φ60×60 | 1.60E7 | 6.72E6 |
| 3 | #78×62 | #76×76 | #90×90 | #84×82 | #90×90 | 1.80E7 | 8.57E6 |
| 4 | Ø12 | Ø15 | Ø58 | φ18×18 | φ30×28 | 1.67E6 | 6.39E5 |
| 5 | Ø12 | Ø15 | #90×90 | φ18×18 | φ30×28 | 1.67E6 | 6.40E5 |
| 6 | Ø12 | Ø50 | Ø40 | φ58×36 | φ42×42 | 1.92E6 | 6.36E5 |
| 7 | Ø12 | Ø50 | Ø58 | φ58×38 | φ62×62 | 1.94E6 | 6.36E5 |
| 8 | Ø12 | Ø50 | #90×90 | φ58×40 | #98×68 | 1.94E6 | 6.42E5 |
| 9 | Ø12 | #76×76 | Ø40 | #66×36 | φ42×42 | 1.92E6 | 6.42E5 |
| 10 | Ø12 | #76×76 | Ø58 | #70×36 | φ62×62 | 1.94E6 | 6.37E5 |
| 11 | Ø12 | #76×76 | #90×90 | #70×38 | #98×68 | 1.98E6 | 6.42E5 |
| 12 | Ø50 | Ø15 | Ø40 | φ14×14 | φ42×26 | 1.04E7 | 2.41E6 |
| 13 | Ø50 | Ø15 | Ø58 | φ14×14 | φ46×26 | 1.04E7 | 2.42E6 |
| 14 | Ø50 | Ø15 | #90×90 | φ14×14 | φ46×26 | 1.04E7 | 2.42E6 |
| 15 | Ø50 | Ø50 | Ø40 | φ54×54 | φ42×42 | 1.60E7 | 6.72E6 |
| 16 | Ø50 | Ø50 | #90×90 | φ54×54 | #88×74 | 1.60E7 | 6.72E6 |
| 17 | Ø50 | #76×76 | Ø40 | #86×76 | φ42×42 | 1.60E7 | 6.72E6 |
| 18 | Ø50 | #76×76 | Ø58 | #86×74 | φ62×62 | 1.60E7 | 6.71E6 |
| 29 | Ø50 | #76×76 | #90×90 | #86×74 | #98×94 | 1.60E7 | 6.72E6 |
| 20 | #78×62 | Ø15 | Ø40 | φ14×14 | φ46×26 | 1.09E7 | 2.42E6 |
| 21 | #78×62 | Ø15 | Ø58 | φ14×14 | φ46×26 | 1.09E7 | 2.42E6 |
| 22 | #78×62 | Ø15 | #90×90 | φ14×14 | φ46×26 | 1.09E7 | 2.41E6 |
| 23 | #78×62 | Ø50 | Ø40 | φ54×54 | φ42×42 | 1.80E7 | 8.57E6 |
| 24 | #78×62 | Ø50 | Ø58 | φ54×54 | φ62×62 | 1.80E7 | 8.58E6 |
| 25 | #78×62 | Ø50 | #90×90 | φ54×54 | #82×70 | 1.80E7 | 8.57E6 |
| 26 | #78×62 | #76×76 | Ø40 | #82×82 | φ42×42 | 1.80E7 | 8.58E6 |
| 27 | #78×62 | #76×76 | Ø58 | #84×82 | φ62×60 | 1.80E7 | 8.57E6 |

The neutron energy spectra at ES#1 and ES#2 for different combinations are also given in Fig. 3 and Fig. 4. The beam energy spectra are slightly affected by the aperture size as a whole. For the energy spectrum at ES#1, around the eV energy region, the neutron energy spectra are slightly higher in the case of large apertures. Because the ES#2 is farther away from the target, the difference in the energy spectrum of ES#2 with different apertures is smaller. Further comparing the energy spectra of the two endstations, as shown in Figures 5 and 6, the energy spectra of the two endstations are significantly different for the two sets of nominal collimation parameters. As the neutron energy increases, the ratio of neutrons gradually decreases. Therefore, the energy spectra of the two endstations must be accurately and independently measured and cannot be simply substituted for each other.

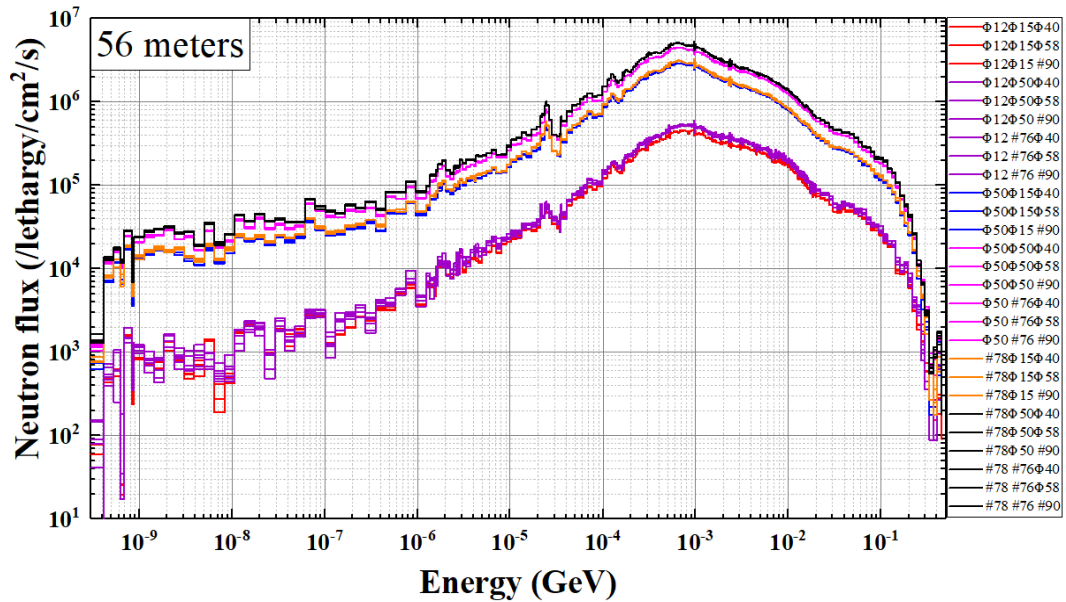

Fig. 3: Calculated and measured energy spectra with different collimating combinations at ES#1.

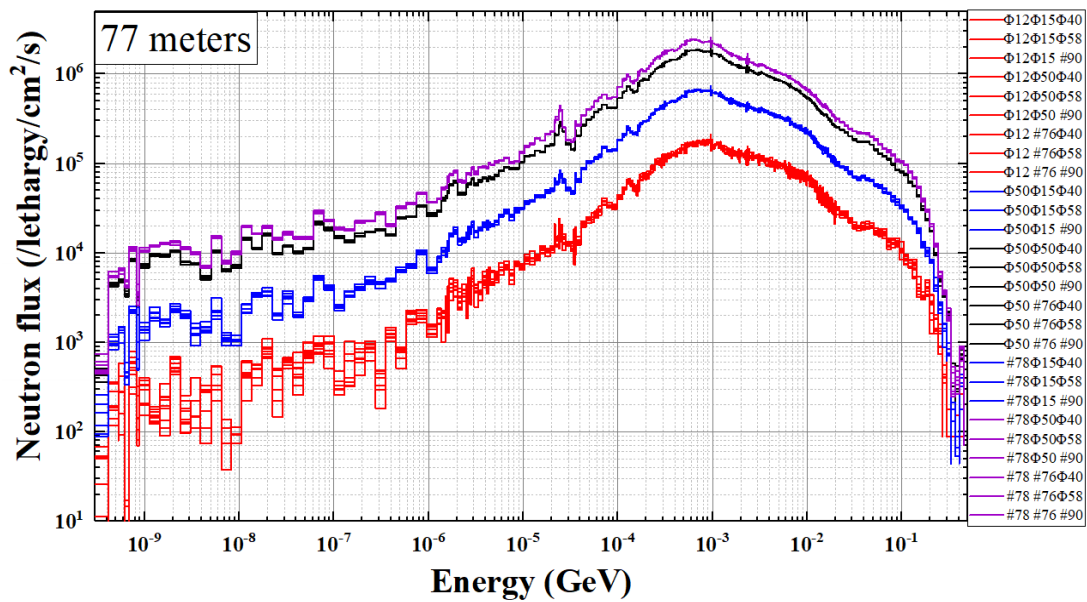

Fig. 4: Calculated and measured energy spectra with different collimating combinations at ES#2.

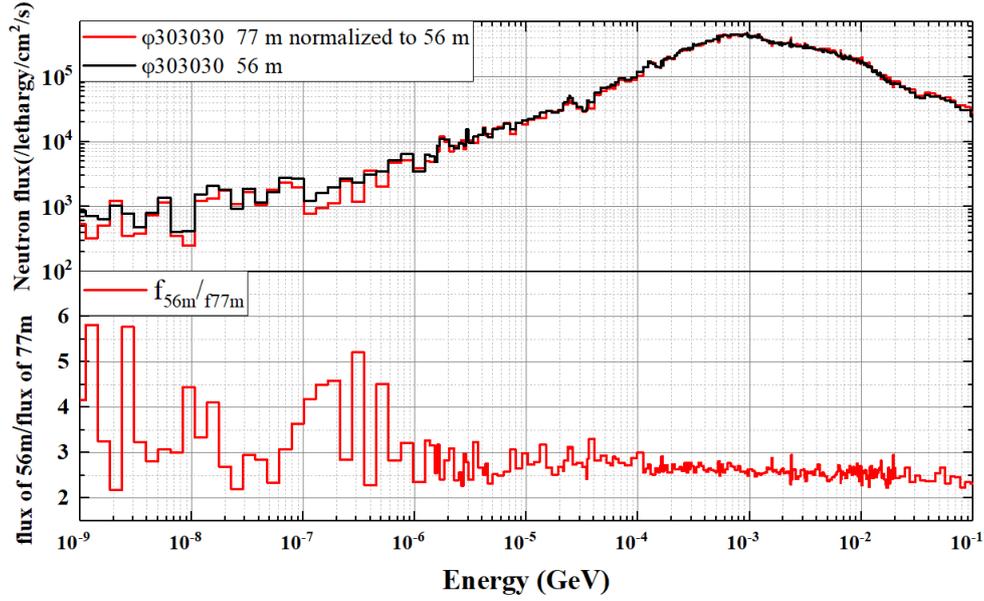

Fig. 5: Beam fluxes and ratio of fluxes at two endstations for a set of standard aperture combinations of the Φ30 mm beam spot.

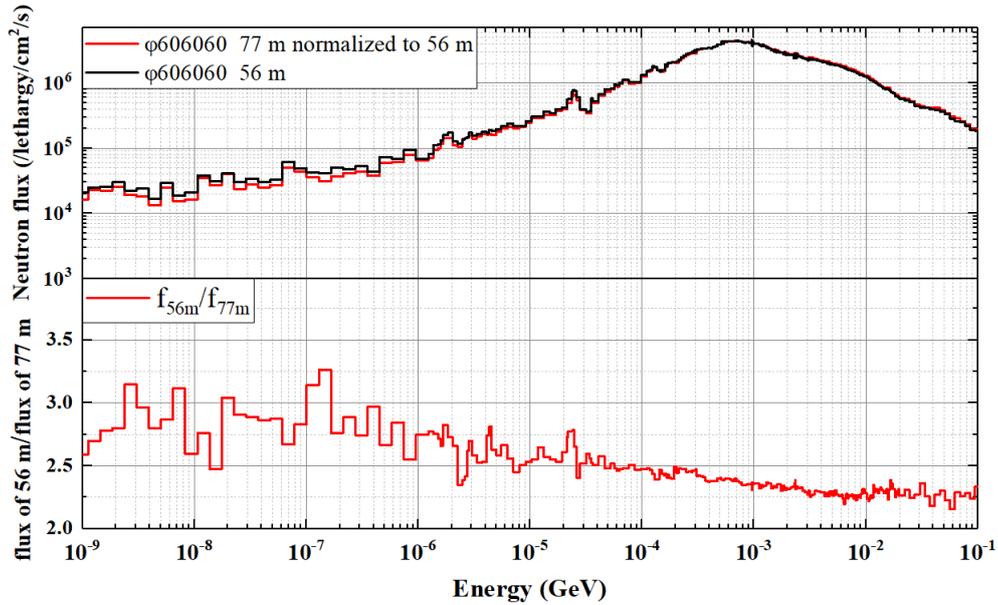

Fig. 6: Beam fluxes and ratio of fluxes at two endstations for a set of standard aperture combinations of the Φ60 mm beam spot.

## 4. Influence of the proton beam spot distribution on the Back-n beam

Since the Back-n neutron beam is guided from the incident target surface of the proton beam, it is more susceptible to the target proton beam spot distribution. There were three-phase changes of proton beam spot distribution on the CSNS target: 1) Before the CSNS facility was built, according to the preliminary physical design, the distribution on the target was set to be rectangular uniform; 2) During commissioning, the beam power was ramped up to 80 kW, the distribution was approximately Gaussian; 3) After the proton beam power of 80 kW, the magnetic lattice for the beam spot homogenization before the target is switched on, and the beam spot is of a uniform rectangle distribution from the measurement of a multi-wire beam monitor.

## 4.1 Fitting of the measured uniform rectangular proton beam spot

The CSNS accelerator currently runs at the full power of 100 kW, and the beam spot homogenization technique based on octopole magnets is used before the target [19]. A rectangular uniform proton beam distribution can be obtained. A multi-wire beam monitor at 1.89 meters from the CSNS target [20] was arranged, which can implement real-time direct measurement of the proton beam profile. Figure 7 shows the horizontal (X) and vertical (Y) profile curves respectively detected by the multi-wire monitor.

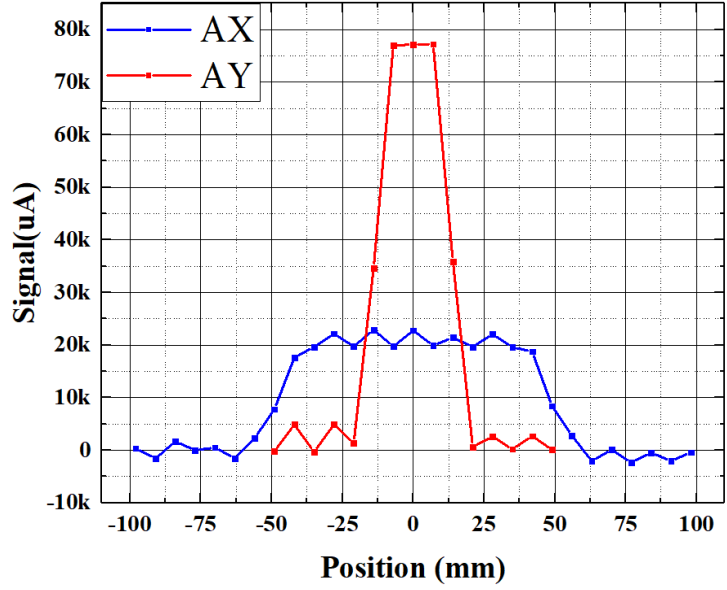

Fig. 7: Signals of proton beam detected by the multi-wire monitor before the target.

According to the distribution characteristics of the curves in Figure 7, a flat top function based on the error function is used for fitting:

$$f_R(x,y) = N \times \left\{\left[erf\left(\frac{\alpha_1+x}{\sqrt{2}\sigma_1}\right) + erf\left(\frac{\alpha_1-x}{\sqrt{2}\sigma_1}\right)\right] \times \left[erf\left(\frac{\alpha_2+y}{\sqrt{2}\sigma_2}\right) + erf\left(\frac{\alpha_2-y}{\sqrt{2}\sigma_2}\right)\right]\right\}, \quad (1)$$

where N is the normalization constant, σ denotes the shape of rising and falling edges, α denotes the width of the curves. erf () is an error function and is written as,

$$\text{erf}(u) = \frac{2}{\sqrt{\pi}}\int_0^u e^{-t^2} dt. \quad (2)$$

We fit X and Y curves in Figure 7 respectively using the formula (1) with a set of parameters, N=1/10368; $\alpha_1$=48 mm; $\alpha_2$=13.5 mm; $\sigma_1$=6.223 mm; $\sigma_2$=2.829 mm, and the results are shown in Figure 8. Furthermore, the 2D contour distribution and 3D distribution are also plotted in Figures 9 and 10.

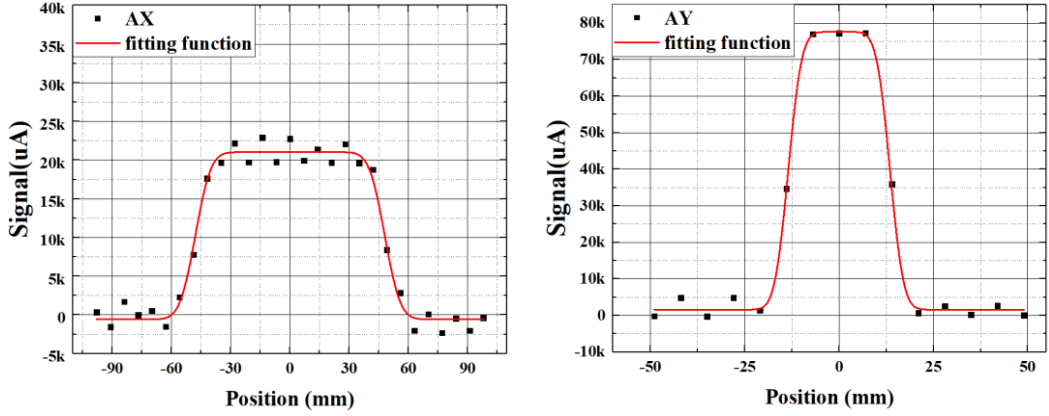

Fig. 8: Measured points from the multi-wire monitor are fitted for X and Y directions.

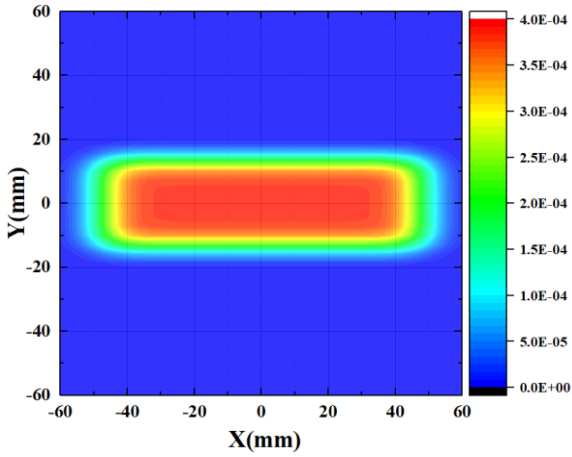

Fig. 9: 2D proton beam spot distribution according to the fitting function.

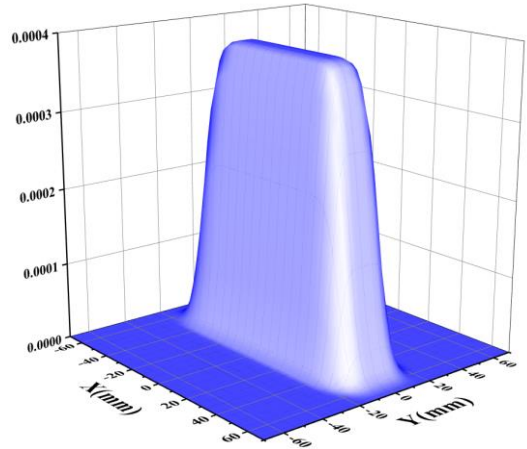

Fig. 10: 3D distribution according to the fitting function.

### 4.2 Fluxes and energy spectra of neutron beam for different proton beam spots

We have analyzed the fluxes and energy spectra of the Back-n white neutron beam for three kinds of distributions of the beam spot, namely the ideal rectangular uniform distribution, Gaussian distribution, and the actual rectangular uniform distribution by fitting measured data using formula (1). The size of the ideal rectangular uniform beam spot on the target is 120 mm×40 mm. The double Gaussian distribution adopted the parameters of $FWHM_X$=88 mm and $FWHM_Y$=33 mm during CSNS commissioning is,

$$f_{gauss}(x,y) = \frac{1}{2\pi\sigma_x\sigma_y} e^{-\frac{1}{2}\left[\left(\frac{x}{\sigma_x}\right)^2 + \left(\frac{y}{\sigma_y}\right)^2\right]} \quad . \tag{3}$$

The simulated fluxes of the Back-n white neutron beamline for three operation modes of proton beam spots are listed in Table 3. The energy spectra of the beamline at 24 meters from the target are shown in Figure 11. There is no difference in the energy spectra obtained with the three proton beam spot modes. The uncertainty of the proton beam spot distribution has a relatively large impact on the white neutron fluxes. As the distance from the target is farther, the flux difference becomes greater. As the distance from the target surface is farther, the fraction of neutrons produced in the center of the target becomes higher and higher. The central distribution of the proton beam spot determines the yield of these neutrons.

Table 3: The flux comparison of different proton beam spot distributions at three positions on the Back-n neutron beamline. The last column is the relative error comparing with the Gaussian distribution.

| Proton distribution | Flux (n/cm$^2$/s) | Position of Detector | Difference |
|---|---|---|---|
| Gaussian | 1.42E8 | Z=-24m (24m wall) | -- |
| Uniform | 1.50E8 | Z=-24m (24m wall) | 5.63% |
| Actual uniform | 1.52E8 | Z=-24m (24m wall) | 7.04% |
| Gaussian | 1.60E7 | Z=-56m (ES#1) | -- |
| Uniform | 1.71E7 | Z=-56m (ES#1) | 6.88% |
| Actual uniform | 1.82E7 | Z=-56m (ES#1) | 10.63% |
| Gaussian | 6.72E6 | Z=-77m (ES#2) | -- |
| Uniform | 7.13E6 | Z=-77m (ES#2) | 6.10% |
| Actual uniform | 7.65E6 | Z=-77m (ES#2) | 12.80% |

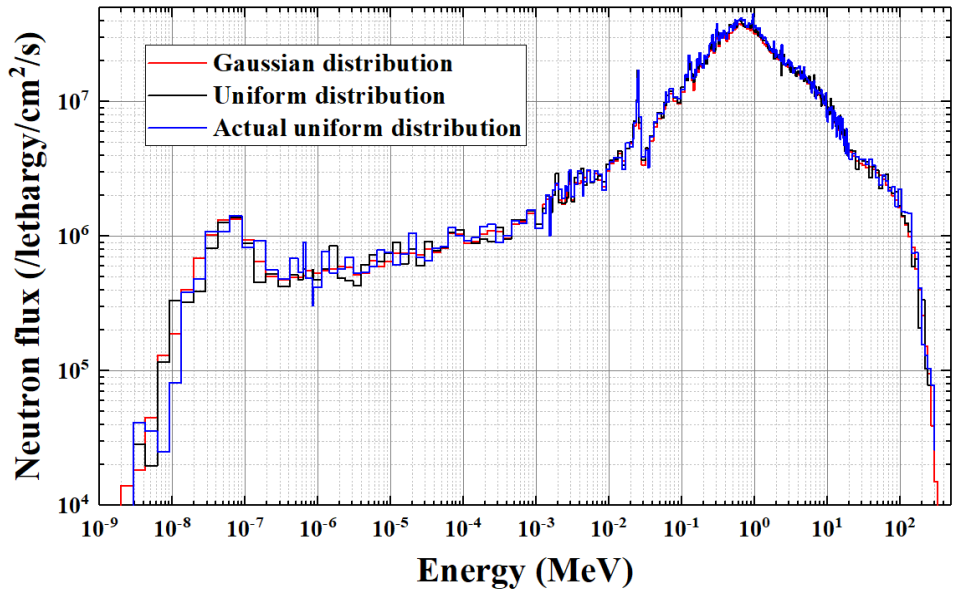

Fig. 11：The energy spectra of the Back-n white neutron beam at 24 meters for the three proton beam spot distributions.

## 5. Conclusion

According to the user demands, we have studied the possibility of extension of the neutron beam operation modes. 27 combinations of apertures on the three collimators are used to simulate neutron transport on the beamline by FLUKA. It is found that when the experiments do not require a high-precision beam condition, it is possible to increase the neutron flux at the endstations by increasing the aperture of the first collimator. However, some combinations of collimating apertures will cause elliptical beam spots and larger inhomogeneity of the beam spot.

For the CSNS proton beam spots at different operation phases, the differences in the fluxes and energy spectra of the Back-n white neutron beam are studied. The difference in the white neutron energy spectra at endstations can be negligible. The fluxes for the actual uniform proton beam spot and Gaussian proton beam spot at ES#1 and ES#2 respectively are quite different, which are 10.63% and 12.80%. Therefore, when the accelerator changes the operation parameters of the proton beam, the Back-n white neutron beam should be calibrated on time.

**Acknowledgements**

This work was supported by the National Natural Science Foundation of China (Project: 12075135) and Guangdong Basic and Applied Research Foundation (No.2018A0303130030); we'd like to thank the colleagues of the CSNS white neutron source collaboration for discussions.